\documentclass[10pt,twocolumn,oneside,notitlepage,final]{revtex4}
\usepackage{color}
\usepackage{epsfig}
\usepackage{amsbsy} %this is the eps fig macro
\usepackage{amsmath}

\begin{document}
%\include{graphicx} %this is the direct pdf macro

%Theory of Condensed Matter Group, Cavendish Laboratory, Madingley
%Road, Cambridge CB3 0HE, United Kingdom

\title{ Ferromagnetic resonance with a magnetic Josephson junction}

\date{\today}

\begin{abstract} We show experimentally and theoretically that there
is a  coupling via the Aharonov--Bohm phase between the order
parameter of a
ferromagnet and a singlet, $s$-wave, Josephson supercurrent. We have
investigated the possibility of measuring the dispersion of such spin 
waves by
varying the magnetic field applied in the plane of the junction and
demonstrated the electromagnetic nature of the coupling by the
observation of
magnetic resonance side-bands to microwave induced Shapiro steps.
\end{abstract}

%\maketitle

%\title{Supplementary Information}

\author{ S. E. Barnes$^{1}$, M. Aprili$^{2}$, I. Petkovi\'{c}$^{3}$,
and S. Maekawa$^{4}$}

\affiliation{$^{1}$Physics Department, University of Miami, Coral
Gables, 33124 FL, USA.  $^{2}$Laboratoire de Physique des Solides, B\^{a}t. 510,
Universit\'{e} Paris-Sud, 91405 Orsay Cedex, France. $^{3}$Service de
Physique de l`Etat Condens\'{e}/IRAMIS/DSM (CNRS URA 2464), CEA
Saclay, F-91191 Gif-sur-Yvette, France. $^{4}$Advanced Science
Research Center, Japan Atomic Energy Agency, Tokai,
Ibaraki 319-1195, and CREST, Japan Science and Technology Agency,
Tokyo 102-0075, Japan.}

\maketitle

Rotation symmetry associated with the O(3) orthogonal group forbids a
coupling between a scalar \textit{s}-wave order parameter and the vector order
parameter $\vec M$ of  the ferromagnet. However a Josephson junction
\cite{barone} defines a plane, the O(3) symmetry is broken, and such a
coupling is
possible. Here we describe a part of the rich spectroscopic magnetic
resonance possibilities that this observation implies. It is possible to
perform a ``photon free" FMR experiment\cite{PRB} on about $10^7$ Ni
atoms, something  infeasible with standard FMR techniques.

Interactions in nature reflect certain gauge groups and the associate
phases which generate a vector potential $\vec A$, called the Berry
connection\cite{Berry}. Interactions via electromagnetic fields
generated by the conserved electrical currents reflect the U(1) gauge
group and the
Aharonov--Bohm (AB) phase\cite{AB}. Associated with angular moment is
the gauge symmetry SU(2) and the familiar Lie algebra, i.e., the spin
commutation rules. The AB is replaced by the spin Berry phase. It is
often imagined that magnetic moments might interact with the
Josephson current by direct spin-flips, which would involve the spin
Berry phase\cite{spin}.  In this paper it will be shown that
quantitatively the
interaction between the Josephson current and the order parameter in
superconductor/ferromagnet/superconductor (SFS) junctions can be
explained in terms of the AB-phase and regular
electrodynamics. It will be shown such an experiment measures rather
directly the magnet correlation function.

\begin{figure}[b]
\vspace{2mm}
\centerline{\hbox{  \epsfig{figure=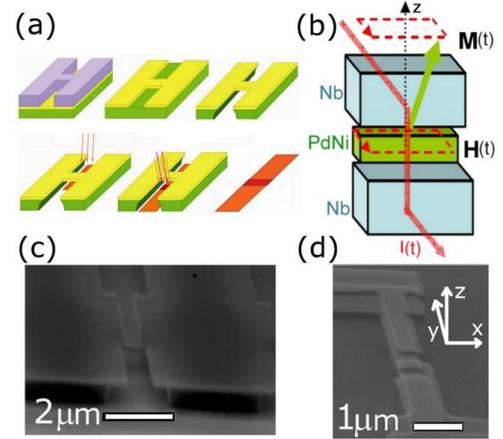,width=70mm}}}
\caption{Description of the sample. \textbf{(a)} The fabrication
procedure. Trilayer mask is etched to obtain the suspended bridge and the undercut, then the junction is fabricated by shadow evaporation and liftoff.
\textbf{(b)} The principle of the experiment. AC Josephson current across the junction excites the spin wave modes which in turn couple to the Josephson current and rectification takes place. \textbf{(c)} Scanning
Electron Microscopy (SEM) picture of the $\rm Si_3N_4$/PES mask before
angle evaporation.\textbf{ (d)} SEM picture of a junction after lift
off. }
\label{Fig1} \end{figure}

We  fabricated $\rm Nb/Pd_{0.9}Ni_{0.1}/Nb$ Josephson junctions by
\textit{in-situ} angle evaporation through a resist mask and
subsequent lift-off.
The mask is defined by e-beam lithography on a Polyether Sulphone - PES
(500 nm)/$\rm Si_3N_4(60\,nm)$/Polymethyl Methacrylate - PMMA (350 nm)
trilayer
\cite{dubos} , and etched in a Reactive Ion Etching (RIE) chamber. The
$\rm Si_3N_4$ is etched  1 minute 30 seconds in $\rm SF_6$ and the PES
in
Oxygen plasma for 10 minutes, giving an undercut of $500$ nm. The mask
fabrication process is schematically presented in Fig.~\ref{Fig1}a.
The $\rm
Si_3N_4$ suspended bridge allows shadow evaporation. A scanning-
electron-microscope (SEM) picture of the mask including the suspended
bridge is
reported in Fig.~\ref{Fig1}c.  The first Nb layer is evaporated at -45
degrees with respect to $z$ axis while the PdNi is evaporated at 45 degrees
and the second Nb layer at 47 degrees. The overlap in the \textit{y}
direction defines the junction area. The shadow
evaporation is illustrated in  the lower drawing array of Fig.~
\ref{Fig1}a. Eight junctions are evaporated on the sample chip. A SEM
image of one of
the junctions after lift-off is shown in  Fig.~\ref{Fig1}d. The
electron-gun evaporation is carried out in ultra-high-vacuum (UHV)
with a base
pressure lower than $10^{-9}$ mbar. The two Nb superconducting films
are $50$ nm thick and the ferromagnetic layer of $\rm
Pd_{0.9}Ni_{0.1}$ is $20$
nm thick. The Ni concentration is measured by Rutherford
Backscattering (RBS) on a test sample. The magnetization loops
obtained by SQUID magnetometry
with in-plane and out-of-plane magnetic field  show a predominant
perpendicular anisotropy. Finally, a schematic view of the junction
including the
principle of the experiment is to be found in Fig.~\ref{Fig1}b.

\begin{figure}[b] \vspace{2mm}
\centerline{\hbox{  \epsfig{figure=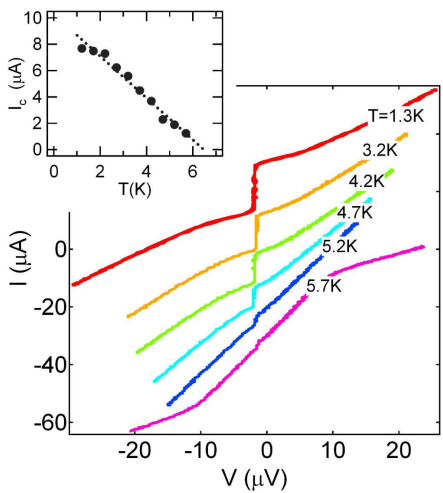,width=70mm}}}
\caption{Current-voltage characteristics with increasing
temperature. The curves have been shifted vertically for clarity. They
do not show any hysteresis as expected when the phase dynamics is
strongly
damped. The critical current as a function of the temperature is shown
in the insert, the dashed line is a linear fit.} \label{Fig2}
\end{figure}

Typical current-voltage ($IV$) characteristics are shown in Fig.~
\ref{Fig2}  as function of temperature. Well below the critical
temperature, the
critical current $I_{c} \approx 7\,\mu$A and normal resistance
$R_n=1\,\Omega$ give a Josephson coupling of $7\,\mu$V, consistent
with early studies
on highly underdamped PdNi-based Josephson junctions  \cite{kontos}.
The $IV$ characteristics are not hysteretic confirming overdamped
phase dynamics
and are well described by the resistively-shunted-junction (RSJ)
model  \cite{barone}. The critical current versus temperature (see
inset of
Fig.~\ref{Fig2}) shows the typical linear behavior expected when the
Thouless energy of the ferromagnetic layer is larger than the Nb
superconducting
energy gap.  This linear dependence has been observed previously in
highly underdamped junctions\cite{kontos}. The junction critical
temperature is
about 7.0 K while the critical temperature of the Nb leads is $7.6$ K
and their critical current over $500\,\mu$A at low temperature. We
have measured
four ferromagnetic junctions on the same wafer, the dispersion of the
critical current from junction to junction is about $\pm$1$\,\mu$A, $
\Delta R_n$
is 0.15 $\Omega$ while the $I_c R_n$ varies by less than 3$\%$ from
junction to junction. We have also  fabricated non-magnetic junctions
by the same
process but replacing the PdNi with a thicker $70$ nm Pd layer. These
junctions have a much larger critical current of about $44\,\mu$A.

For a square junction of side $L$, the total super-current is given
by  the integral \cite{barone}

\begin{equation} %\label{current}
I_s = J_c
\int_{-L/2}^{L/2} dx\int_{-L/2}^{L/2} dy \sin \phi(x,y,t) \label{une1}
\end{equation}

\noindent with \vspace{-5pt}

\begin{equation} %\label{phase}
\phi(\vec{r},t)  = \phi_0 +  \omega_J t  - \frac{2e}{\hbar} \int
\vec{A}\cdot d \vec{r}, \label{une2}
\end{equation}

\noindent where $\phi_0 $ is an
arbitrary phase, $\omega_J = (2e/\hbar) V_0$ is the Josephson
frequency, and the last term  is the AB phase \cite{rowell}, involving
the vector
potential $ \vec{A}$. We use a gauge where $\vec{A} =
A(\vec{r},t)\, \mathbf{\hat{ z}}$, the direction $\mathbf{\hat{z}}$
being perpendicular
to the junction surface. Therefore $\phi(\vec{r},t) = \phi_0 + kx +
\omega_J t -\phi_1$, where $\phi_1 $
%= (4a e/\hbar) A_{mz}$
reflects time
dependent fields and $k = (4ed/\hbar) \mu_0 H + (4ea/\hbar) \mu_0
M_{0y}$.  Here $M_{0y}$ is the $y$-component of the static
magnetisation $
\vec{M_0}$, the applied field $\vec{H}$ is in the $y$ direction,
$2a$ and $2d\!=\!2(a+\lambda)$ are the actual and magnetic thickness
of barrier
and $\lambda$ the London penetration depth. These equations describe
both the statics and the dynamics of our junctions.

The approach is similar to that used for junctions with magnetic
impurities in a normal metal barrier\cite{oldseb}. It is necessary to
determine the
appropriate boundary conditions for the solutions of Maxwell's
equations. For  reasons of transparency, it is not at all useful to
solve the very
difficult problem in which  the solution within the barrier is matched
to that in the exterior region to the junction. Within the junction we
can
ignore the displacement and transport currents since the wavelength of
light $\lambda$ and the skin depth $\delta$ are both larger than the
dimensions
of the junction at the Josephson frequency $\omega_J$ relevant for the
FMR. We observe that the impedance of the junction of $\sim1\,\Omega$
is much
smaller than 377 $\Omega$ of free space and as a consequence there is
essentially no radiation from the junction. The displacement current,
and
evidently, the transport current can therefore also be ignored in the
exterior region. It is therefore only necessary to integrate Amp\`{e}re's
circulation law

\begin{equation} \label{200 }
\vec \nabla \times \vec H = J_s(\vec r,t) \hat {\bf z}; \ \ \   J_s =
J_c \sin \phi(\vec r,t)
\end{equation}

\noindent where $\phi_0$ is absorbed by a time translation, and $
\phi(\vec r,t) = kx + \omega_J t - \phi_1$. Required is the additional
AB phase
shift

\begin{equation} \label{S201}
\phi_1 = \frac{2e}{\hbar}\int_{\vec r}^{\vec r+ 2a \hat{\bf z}} \vec
A^1 \cdot d \vec r =  \frac{4ae}{\hbar}A^1_z,
\end{equation}

\noindent with the magnetic system reflected in $\vec A^1$. In linear
response  $\phi_1$ is considered as a perturbation and the dc signal is

\begin{equation} \label{S201}
I_1 =  - \frac{4ae}{\hbar  \omega_J} \frac{1}{T} \int_0^Tdt \int_{-L/
2}^{L/2}dx  \int_{-L/2}^{L/2} dy \frac{\partial
J_s}{\partial t} A^1_z
\end{equation}

\noindent which includes a time average over a single period $T$. The
determination of $\vec A^1$ requires first the
vector integration of $\vec \nabla \times \vec H = J_s(\vec r,t)\hat
{\bf z}$ and then $\vec \nabla \times \vec A = \vec B$ with $\vec B =
\mu_0 (\vec
H + \vec M)$. Even with the simplifications of the previous paragraph,
this is an involved calculation. It is useful to make some formal
manipulations
in order to avoid this double integration. First $I_1$ is written as

\begin{equation} \label{S210} I_1 = -  \frac{4ae}{\hbar \omega_J}
\frac{1}{T}
\int_0^Tdt  \int_{-L/2}^{L/2}dx  \int_{-L/2}^{L/2} dy
\left( \frac{ \partial }{\partial t}   \vec \nabla \times \vec H
\right) \cdot \vec A
\end{equation}

\noindent using Amp\`{e}re's law $\vec J = \vec \nabla \times \vec H$.
Performing an integration by parts on time we have

\begin{equation} \label{S210}
I_1 =  \frac{4ae}{\hbar \omega_J}\frac{1}{T} \int_0^Tdt  \int_{-L/
2}^{L/2}dx  \int_{-L/2}^{L/2} dy \left(\vec \nabla \times \vec H
\right) \cdot
\frac{ \partial \vec A}{\partial t}.
\end{equation}

\noindent Using the fact that the Poynting vector, and hence $\vec H
\times { \partial \vec A}/{\partial t}
= 0$ on the surface, this is integrated again by parts using $\vec
\nabla \cdot(\vec H\times (\partial \vec A/\partial t)) =  (\partial
\vec
A/\partial t) \cdot \vec \nabla \times \vec H- \vec H \cdot \vec
\nabla \times (\partial \vec A/\partial t) $, to give

\begin{equation} \label{S212}
I_1 = \frac{4ae}{\hbar \omega_J} \frac{1}{T} \int_0^T \int_{-L/2}^{L/
2}dx  \int_{-L/2}^{L/2} dy \vec H \cdot  \frac{\partial }{\partial t}
\vec
\nabla \times \vec A.
\end{equation}

\noindent Then, since $\vec \nabla \times \vec A= \vec B$ and $\vec B
= \mu_0 (\vec H + \vec M)$,  the signal

\begin{equation} \label{S220}
I_1 = - \frac{1}{V_0}\int dv \overline{ \vec H \cdot  \frac{\partial
\vec M}{\partial t} }
\end{equation}

\noindent where $dv = 2a
dx dy$ is the elementary volume, the integral is over the volume of
the magnetic layer, and the average is  indicated by the bar. The
resonance
of the ferromagnetic layer is contained in $\chi_i(t) $, the dynamic
susceptibility, and

\begin{equation} \label{S221}
M_i (\vec{ r}, t) =
\int d \vec{ r}^\prime \int dt^\prime \chi_i(\vec{ r}-
\vec{ r}^\prime; t-t^\prime) H_i(\vec{ r}^\prime,t^\prime),
\end{equation}

\noindent where $i = x,y,z$. The simple expressions Eqs.~(\ref{S220})
and (\ref{S221}) are a principal theoretical result presented here and
have an obvious
interpretation in terms of the magnetic energy. They demonstrate that
Josephson junction magnetic spectroscopy measures very directly the
magnetic
susceptibility correlation  function $ \chi_i(\vec{ r}- \vec{ r}^
\prime; t-t^\prime)$, and all the other excitations to which that
couples,  in
much the same manner as does neutron scattering. As will be seen
below, the advantage is that this technique couples preferentially to
small $q$ wave
vectors.

Since $H_i(\vec{ r},t^\prime)$ is periodic in time, the {\it time
\/} convolution Eq.~(\ref{S221}), reduces to a product and {\it if\/}
the
susceptibility is sufficiently local and then

\begin{equation} \label{S222}
M_i (\vec{ r},\omega_J) = ( \chi^\prime_i(\omega_J) + i
\chi^{\prime\prime}_i(\omega_J) ) H_i(\vec{ r},\omega_J)
\end{equation}

\noindent in the usual complex notion.  If rather the response is non-
local then

\begin{equation} \label{S223}
M_i (\vec{ k},\omega_J) = ( \chi^\prime_i(\vec{ k}, \omega_J) +
i \chi^{\prime\prime}_i(\vec{ k},\omega_J) )
H_i(\vec{ k},\omega_J)
\end{equation}

\noindent and it requires a Fourier expansion of the spatially
dependent $H_i(\vec{ r},\omega_J)$. Finally, when
$\omega_J \approx I_c R$, as for the lowest voltage experimental
signals, the linear response approximation is not strictly applicable
and high
harmonics of $\omega_J$ must be accounted for. Similar expressions
apply but now, in particular, ``half-harmonic'' signals occur since the
super-current contains {\it higher harmonics\/} and can, corresponding
to the second harmonic, excite a resonance at $\omega_s$ when $
\omega_J /2 =
\omega_s$.

Now {\it all\/} that is required is  a {\it single\/} integration of

\begin{equation} \label{S224}
\vec \nabla \times \vec H  = J_c \sin [kx +
\omega_J t] \hat{\bf z} ,
\end{equation}

\noindent but which is not a simple task. It is trivial to verify by
differentiation that such an integral is $\vec H_p
= -(J_c/k) \cos (kx + \omega_J t)\hat{\bf y}$. However this does not
satisfy the boundary condition that the current density is zero
outside the
square junction region. With the present gauge $\vec A = A_z {\hat{\bf
z}}$, it is necessary to find a solution of Laplace's equation  $
\nabla^2 A_z
=0$ such that $\vec H =  \vec H_p + \vec H_i$ where $\vec H_i = \vec
\nabla \times \vec A $ is such that $\vec H$ satisfies
Eq.~(\ref{S224}) inside
the square but has $\vec \nabla \times \vec H=0$ outside. A little
reflection suggests there are two contributions to $\vec H_i $ which
must be
accounted for. First, in general, reflecting the even part $J_c  \cos
kx \sin \omega_J t$ of the current density there is a net oscillating
super-current $I_c(H)$ which causes a {\it circulating} magnetic field about $
\hat{\bf z} $, and second, associated with the spatially odd
part of $J_c
\sin [kx + \omega_J t]$, there is a uniform component of the field in the $
\hat{\bf y}$-direction. The first contribution is determined by
considering the
problem with $k=0$, i.e., with a uniform current density $
\overline{J}_ s$. The solution is $ \vec H_s = \frac{1}{2}
\overline{J}_ s (x\hat{\bf y} -
y\hat{\bf x}) \cos \omega_J t  +  \ldots $ where the ellipsis reflects
the relatively small corrections for a square as compared with a
circular cross
section. In what follows this correction is ignored. The corresponding
vector potential has $A_z = (1/4)\overline{J}_ s(x^2+ y^2)$. For
finite $H$,
integrating the even part of the current density gives an average
super-current density of $\overline{J}_ s =  J_c%\frac{\sin \frac{k L}{2}}{\frac{k L}{2}}.
\sin (k L/2)/(k L/2).$
By inspection it is observed $A_z = (1/4)\overline{J}_ s(-x^2+ y^2)$
satisfies $\nabla^2 A_z =0$. The
corresponding odd $ \vec H_i^o = %\frac{1}{2} (1/2) \overline{J}_ s (-
x\hat{\bf y} - y\hat{\bf x}) \cos \omega_J t  +  \ldots. $ The sum $
\vec H_p +
\vec H_i^o$ correctly reduces to $\vec H_s $ in the limit $k\to 0$. It
is the case that $\vec H_p$ (and $\vec H_p + \vec H_i^s$) implies a
uniform
oscillating field but one which diverges as $k \to 0$, whereas
physically, the even part of $\vec H_p$ should be proportional to $k$
reflecting the
Josephson screening of  fields. That the tangential applied field $H$
be continuous requires the current, induced by the even part of $\vec H
$, to be
zero at the surface. The even part of $\vec H_p$ is $ - J_c \frac{\cos
kx }{k} \sin \omega_J t \hat{\bf y} ({\cos kx }/{k} )\sin \omega_J t
\hat{\bf
y} $ The required even part of $H_i $  is now $ \vec H_i^e =  J_c
\frac{\cos k\frac{L}{2}}{k} \sin \omega_J t \hat{\bf y} ({\cos k
\frac{L}{2}}/{k})
\sin \omega_J t \hat{\bf y} $, which is equally divergent as $k\to
0$. The net result of integrating Eq.~(\ref{S224})  is therefore

\begin{equation} \label{S240}
\vec H = \frac{J_c}{k} [ \cos (kx + \omega_J t) -  \cos \frac{kL}{2}
\sin \omega_J t ]\hat{\bf y} $$ $$ +  \frac{1}{2}
J_c\frac{\sin \frac{k L}{2}}{ \frac{k L}{2}} (- x\hat{\bf y} - y
\hat{\bf x}) \cos \omega_J t  +  \ldots
\end{equation}

\begin{figure}[b] \vspace{2mm}
\centerline{\hbox{  \epsfig{figure=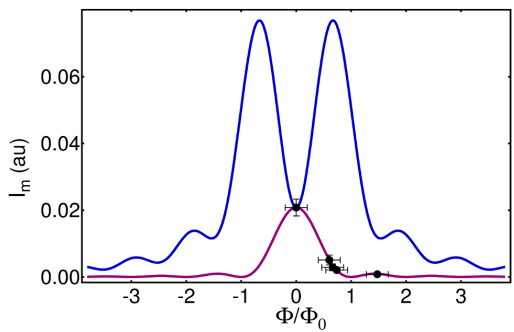,width=70mm}}} \caption{In
blue is $F_y$ and in pink $F_x$. These curves
satisfy the evident requirement that the coefficient of the $x$ and $y
$ responses be the same when $B=0$. Markers denote the experimental values of the
resonance amplitude as function of the applied field, with the normalization constant as the fitting parameter.} \label{Fig3} \end{figure}

Imagine that the magnetic layer is composed of a number of independent
crystallites so that the response is local and Eq.~(\ref{S222}) would
apply.
This local assumption also has the merit of being an useful
illustration of the theory since it leads to a relatively simple
prediction of the $H$
dependence of the signal which can be compared with experiment. This
helps determine if the response is indeed local, or extended. There
are some
complicated integrals involved in the evaluation of Eq.~(\ref{S220}).
The result is written, in closed form, as

\begin{equation} \label{trois4}
I_m =2
\pi I_c (0) \frac{ \Phi_{\rm rf}}{\Phi_0}\left[F_x\chi^{\prime
\prime}_x (\omega_J) + F_y\chi^{\prime\prime}_y (\omega_J)\right],
\end{equation}
\noindent where
$\Phi_{\rm rf} = (2aL) B_{\rm rf} = (2aL)  \mu_0 I_c(0)/ L$ is the
flux due to the radio frequency field and where $$ F_x = \frac{1}{48}
\left[\frac{I_c(B_0)}{I_c}\right]^2 $$
\noindent and
$$
F_y = \frac{2}{x^2}\{1-  \frac{1}{x}\sin \frac{x}{2} \cos\frac{x}{2}
-[\frac{11}{24} +\frac{2}{x^2}] \sin^2 \frac{x}{2}\},
$$ with $x=kL$,  reflect the geometrical structure of the coupling.
The equilibrium magnetization is along the $z$ axis, and the magnetic
resonance
signal is contained in $\chi^{\prime\prime}_x (\omega_J)$ and $
\chi^{\prime\prime}_y (\omega_J)$,  the Fourier transforms of the
imaginary part of the
susceptibility. The two functions $F_x$ and $F_y$ are plotted in Fig.~
\ref{Fig3}. As required by symmetry $F_x=F_y$ when $k=0$. The
$\chi^{\prime\prime}_x (\omega_J)$ response near the first  zero in $
I_c(B_0)=I_c$ is about four times the maximum response to $\chi^{\prime
\prime}_y
(\omega_J)$. While there is a clear reflection of the Fraunhofer
diffraction pattern in the $\chi^{\prime\prime}_y (\omega_J)$
response, the $1/k^2$,
i.e., $1/H^2$ response dominates that to $\chi^{\prime\prime}_x
(\omega_J)$ with only modulations due to diffraction effects. With a
single flux
quantum threading thorough the junction the $F_x$ is zero reflecting
the net absence of a circulating current. In contrast $F_y$ is a
maximum since
the current is odd and the junction constitutes a small flat
solenoid carrying the critical current density and hence a maximum
field internal to
the junction.

Given a static magnetisation $\vec M = M_z \hat{\bf z}$ the magnetic
susceptibility might be approximated by

\begin{equation} \label{S500}
\chi^{\prime\prime}_x(\vec k,\omega) = \chi^{\prime\prime}_y(\vec k,
\omega)  \approx \gamma_e M_z \sum_{\pm} \frac{ ( \frac{1}{\tau})}
{(\omega_s + a
k^2 \mp \omega)^2 }
\end{equation}

\noindent where $\tau$ is the relaxation time, $\omega_s$ the
frequency of the FMR mode and $a$ the spin-wave stiffness. Here
$\gamma_e = \mu_B /\hbar$ with $\mu_B$ the Bohr magneton.

\begin{figure}[b] \vspace{2mm}
\centerline{\hbox{  \epsfig{figure=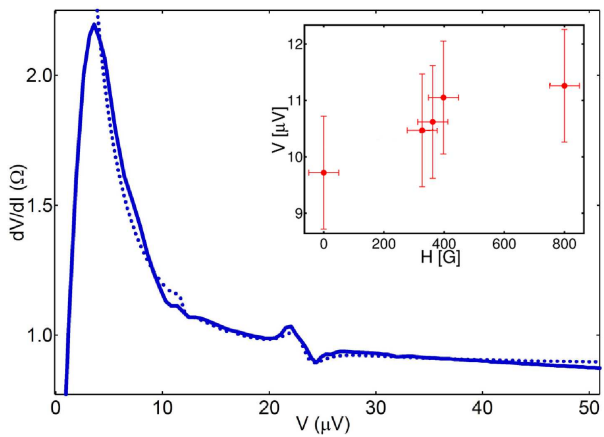,width=85mm}}}
\caption{Dynamical resistance of a ferromagnetic Josephson junction
showing ferromagnetic resonances with first and second harmonic.
Insert: the magnetic field dependence of the resonance at the second
harmonic. The fit takes into account the spatial dispersion of the spin wave mode, see text.}
\label{Fig4} \end{figure}

In Fig.~\ref{Fig4} we report the dynamical conductance at zero applied magnetic field (solid line) and the theoretical expectation (dotted line)  from Eq.~(\ref{trois4}) and
(\ref{S500}) with $a=0$, using $\omega_s$ and  $1/ \tau$  as a parameters. We obtained $\omega_s$=23 $\mu$V as expected from the Kittel's formula $\omega_s = \gamma_e \sqrt{(H_K - 4\pi M_S)^2 - H^2}$, where the
anisotropy field $H_K$ and saturation magnetisation $M_S$ have been measured separately by SQUID magnetometry \cite{PRB}. The FMR frequency is consistent with the value obtained in a reference sample by EPR spectroscopy \cite{PRB}, i.e., the observed signals are entirely consistent with the coupling of the FMR resonance to the superconductivity via the AB-phase. The resonance at  11.5 $\mu$V is a subharmonic of the main mode. Kittel's formula predicts that $\omega_s$ decreases with increasing $H$ whereas the half-harmonic signal of Fig.~\ref{Fig4} actually seems to increase (see inset of Fig.~\ref{Fig4} ). This leads us to believe that this signal corresponds to a finite value of $a$ and $k$ in Eq.~(\ref{S500}). That the spin-wave dispersion might be measured in our type of experiment is an exciting possibility warranting further investigation. Moreover, in Fig.~\ref{Fig3} we also report the amplitude of the FMR signal (markers) as a function of the applied magnetic field. The experimental data follow the  $F_x$  coupling function obtained above.

\begin{figure}[t] \vspace{2mm}
\centerline{\hbox{  \epsfig{figure=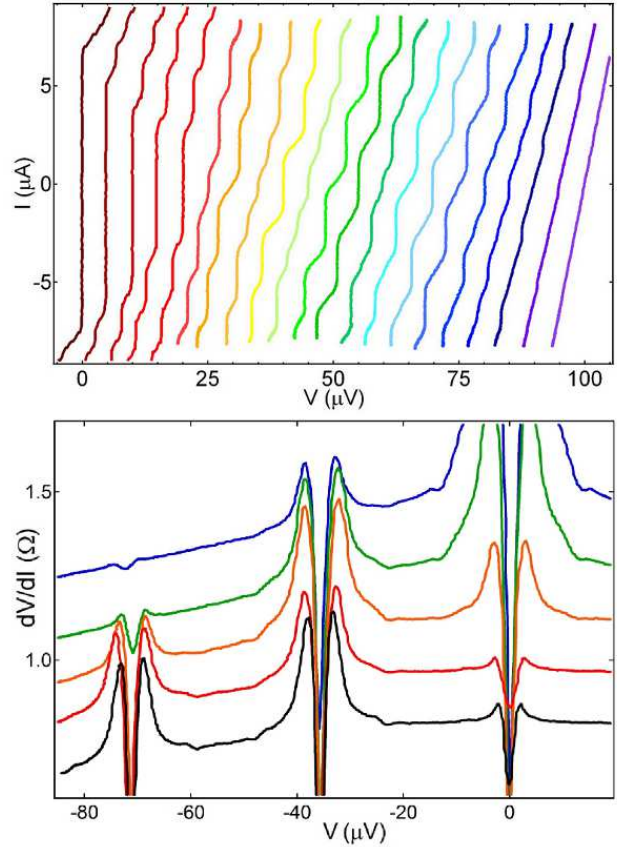,width=85mm}}} \caption{Top
panel: $IV$ curves showing the Shapiro steps, taken at
T=1.3 K and with the microwave frequency of 1 GHz with different
incident power. Bottom panel: the dynamical resistance showing Shapiro
steps and side-band
resonances, taken at T=35 mK and with microwave frequency of 17.35
GHz. In passing from the bottom to top the power increases from 0 dBm
to 20 dBm, in
steps of 5 dBm.} \label{shapiro} \end{figure}

The electromagnetic nature of the response can be confirmed by a study
of Shapiro steps\cite{shapiro,barone}. In the absence of a magnetic
resonance
mode, an applied radio frequency field gives rise to such steps. In
order to account for these  we write for the bias voltage across the
junction
$$ V = V_0 + v \cos \Omega t $$ where  $v$ and $\Omega$  are the
\noindent amplitude and frequency of the applied microwave field while the
constant $V_0$, as
usual, corresponds to a Josephson frequency $\omega_J = 2eV_0/\hbar$.
The Josephson current is now $J_c$ times $$
\sin [kx + \omega t +  \frac{2ev}{\hbar\Omega} \sin \Omega t].
$$ Expanding this sine with the  Jacobi-Anger identity  gives $$
\left[ J_0(\frac{2ev}{\hbar\Omega})S_0 +
\sum_{n=1}^{\infty}\left(J_{2n}(\frac{2ev}{\hbar\Omega}) S_{2n} +
J_{2n-1}(\frac{2ev}{\hbar\Omega})S_{2n-1}\right)\right] $$ with
$S_{2n} =
\sum_\pm \sin(kz + \omega_0t +2n  \Omega t)$ and $S_{2n-1} = \sum_\pm
\pm \sin(kx\pm  \omega_0t +(2n-1)  \Omega t)$, where $S_0 = S_{2n=0}$.
Each term
is of the form

\begin{equation}
J_c^n \sin[kz + \omega_0t \pm n  \Omega t] \to J_c^n \sin[kz +
\omega^n t]
\end{equation}

\noindent where, in well known
fashion\cite{barone}, $J_c^n = J_c J_{n}(2ev/\hbar\Omega) $ involves
the appropriate Bessel function $J_{n}(2ev/\hbar\Omega) $. For the
voltage at
which $\omega^n=0$ there is a direct contribution $J_c^n \sin(\phi_0 +
kx)$ to the average current which is equivalent to the zero voltage
critical
current step but displaced to $V_n = n \hbar \Omega/2e$ and reduced
from $I_c(H) $ to $I_c(H) J_{n}(2ev/\hbar\Omega)$. This corresponds to
the
principal Shapiro steps, shown on the top panel of  Fig.~
\ref{shapiro}.  For the magnetic response, in the linear response
regime, the theory developed
without an applied microwave field can be adopted. All that is needed
is to replace $J_c$ by $J_c^n$ and $\omega_J$ with $\omega^n$ in the
appropriate
places as described above.

We have investigated the Shapiro steps for different microwave power.
The top part of  Fig.~\ref{shapiro} shows the appearance of the
Shapiro steps
in the current-bias characteristics for increasing microwave power.
The amplitude of each step follows the appropriate Bessel function
$J_{n}(2ev/\hbar\Omega) $, as expected. In the bottom panel of   Fig.~
\ref{shapiro} is presented the dependence of the magnetization induced
side-bands
on the micro-wave power. To make these bands more evident, shown is
the dynamical resistance as a function of the  voltage. It is clear
from the data
that the amplitude of the side-band resonances follows the Bessel
function of the Shapiro steps, as expected from the theory described
above. Indeed
the Shapiro steps can be seen as ``replica'' of the critical current
at finite bias and hence the side-band amplitude follows that of the
steps. The
two sidebands correspond to the two poles of the dynamical
susceptibility, Eq.~(\ref{S500}).

In conclusion, we have described experiments and developed theory to
demonstrate that the relative AB phase of the superconductors which
comprise a
Josephson junction couples to the magnetic order parameter of a
ferromagnet. We have thereby performed an FMR experiment with a
sensitivity which
greatly exceeds that of  conventional cavity FMR. Since the coupling
is via the magnetic field it is not necessary to have the current pass
through the
magnetic material. It might be imagined that the magnetic layer be the
top layer of a FSIS structure in which the adjacent S-layer has a
thickness of
the order of, or less than, the London penetration length. The
possibility of measuring the dispersion of spin-wave excitations has
also be
investigated. Our method\cite{PRB,oldseb,flux} of coupling
superconductivity to magnetism measures directly the dynamic
susceptibility
$\chi^{\prime\prime}(\vec q,\omega) $ with an enhanced sensitivity for
small wave-vectors, complementary to neutron scattering.

\end{document}